\documentclass[useAMS,usenatbib]{mn2e} \usepackage{epsfig}
\usepackage{amsmath}
\usepackage{natbib}

\author[Fortenberry and Lukemire] {Ryan C.
Fortenberry\thanks{Email:rfortenberry@georgiasouthern.edu} and Joseph A.
Lukemire \\ Georgia Southern University, Department of Chemistry, Statesboro,
GA 30460 U.S.A.}

\title[C$_3$P$^-$]{Electronic and Rovibrational Quantum Chemical Analysis of
C$_3$P$^-$: The Next Interstellar Anion?}

\begin{document}

\date{Submitted: \today}

\maketitle

\begin{abstract}

C$_3$P$^-$ is analogous to the known interstellar anion C$_3$N$^-$ with
phosphorus replacing the nitrogen in a simple step down the periodic table.  In
this work, it is shown that C$_3$P$^-$ is likely to possess a dipole-bound
excited state.  It has been hypothesized and observationally supported that
dipole-bound excited states are an avenue through which anions could be formed
in the interstellar medium.  Additionally, C$_3$P$^-$ has a valence excited
state that may lead to further stabilization of this molecule, and C$_3$P$^-$
has a larger dipole moment than neutral C$_3$P ($\sim 6$ D vs.~$\sim 4$ D).
As such, C$_3$P$^-$ is probably a more detectable astromolecule than even its
corresponding neutral radical.  Highly-accurate quantum chemical quartic force
fields are also applied to C$_3$P$^-$ and its singly $^{13}$C substituted
isotopologues in order to provide structures, vibrational frequencies, and
spectroscopic constants that may aid in its detection.


\end{abstract}

\section{Introduction}

It is currently uncertain as to why anions are found in the interstellar medium
(ISM) with six known to date: C$_4$H$^-$, C$_6$H$^-$, C$_8$H$^-$, CN$^-$,
C$_3$N$^-$, and C$_5$N$^-$ \citep{Cernicharo07, Brunken07, Remijan07,
Thaddeus08, Cernicharo08, Agundez10, Cordiner11}.  Since all of the known
anions are closed-shell and, as a result, valence in nature, one of the leading
theories about their existence comes from the presence of a dipole-bound
excited state \citep{Agundez08}.  If a neutral molecule possesses a large
enough dipole moment (somewhere in excess of 2.0$-$2.5 D), it can bind an
additional electron in a highly diffuse $s$-type orbital \citep{Fermi47,
Coulson67, Gutsev95, Gutsev95a, Compton96, Jordan03, Simons08, Simons11}.
These loosely bound excited states \citep{Hammer03} can relax to a lower energy
valence state if the molecule can support the additional electron in its
valence electron cloud.  This phenomenon is known for some small anions
\citep{Lykke87, Mullin92, Mullin93}, and its reverse for electronic excitation
has even been hypothesized as a potential explanation \citep{Sarre00,
Cordiner07, Fortenberry13CH2CN-} for at least one of the diffuse interstellar
bands, the series of visible to near-infrared absorption features largely
uniformly observed towards many stellar objects \citep{Sarre06} whose carriers
have only very recently come into focus \citep{Campbell15}.  As such, it is
currently hypothesized that the large dipole moment of the neutral-radical
captures an incident electron whereby the dipole-bound anion then decays down
to the stable, valence ground electronic state \citep{Agundez08,
Fortenberry13C3H-}.

All of the known interstellar anions possess large enough dipole moments to
undergo this process save for CN$^-$ which is likely formed by collisional or
chemical processes \citep{Larsson12, Carelli14}.  The greater
[C$_6$H$^-$]/[C$_6$H] ratio as compared to the significantly smaller
[C$_4$H$^-$]/[C$_4$H] ratio is most elegantly explained by the dipole-bound
formation hypothesis \citep{Agundez08} especially since C$_4$H is more abundant
than C$_6$H, but C$_6$H$^-$ is more abundant than C$_4$H$^-$.  C$_4$H must be
excited into its low-lying $^2\Pi$ excited state in order to have a dipole
moment large enough to capture the additional electron.  C$_6$H is $^2\Pi$ and
strongly dipolar in its ground state meaning that capture of the dipole-bound
electron is much more likely for the longer hydrocarbon chain.  Additionally,
C$_2$H$^-$ has not been detected.  Its corresponding neutral-radical is very
abundant \citep{Heikkila99}, but the dipole moment of the unchanged species is
not large enough to support a dipole-bound state.  Finally, all of the
corresponding neutral-radicals are known to exist in the ISM for each of the
closed-shell anions detected \citep{McCarthy01}.

However, no new anions have been detected in the ISM since 2010 as reported in
the literature \citep{Agundez10}.  Several candidate anions have been studied
quantum chemically in order to provide accurate predictions of their
rovibrational nature.  These include CH$_2$CN$^-$, $c$- and $l$-C$_3$H$^-$, and
CCOH$^-$ \citep{Fortenberry13CH2CN-, Fortenberry13C3H-, Fortenberry14cC3H-,
Fortenberry15CCOH-}.  Another tantalizing potential interstellar anion is
C$_3$P$^-$.  It is analogous with the known C$_3$N$^-$ anion, and several
phosphorus-containing carbon chains (CP, HCP, and CCP) have been detected in
the ISM \citep{Guelin90, Agundez07, Halfen08}.  However, little is known about
$l$-C$_3$P$^-$.  The linear neutral-radical is known to be the lowest energy
isomer on the [C,C,C,P] potential energy surface and has a strongly dipolar
(3.797 D) $^2\Pi$ ground state \citep{delRio96, Maclean08}.  Hence, this
present work sets to provide both electronic as well as high-level
rovibrational data on C$_3$P$^-$ in order to assist in its potential detection.

Quantum chemical studies of excited states of anions have recently shown that
many anions may possess dipole-bound excited states, and some anion excited
states may also be valence in nature \citep{Fortenberry11dbs,
Fortenberry112dbs, Fortenberry133DBS, Fortenberry144DBS, Fortenberry14DIBs,
Theis15}.  Excited states of anions have been experimentally known for
many species \citep{Lykke87, Mullin92, Mullin93, Grutter99, Pachkov01, Pino02,
Pino04}, but the number of dipole-bound excited state candidate anions has
increased with the improvement of computational tools.  Additionally,
high-level rovibrational computations have been able to produce rotational
constants to within 15 MHz of experiment and vibrational frequencies to within
1.0 cm$^{-1}$ of laboratory results in some cases \citep{Huang11, Inostroza11,
Fortenberry12hococat, Fortenberry12HOCScat, Huang13NNOH+, Zhao14,
Fortenberry14C2H3+}.  Consequently, both approaches are applied to C$_3$P$^-$
in this work in order to elucidate more fully the chemical physics of this
potentially astrochemically significant anion.

\section{Computational Details}

All computations make use of coupled cluster theory at either the singles and
doubles (CCSD) or singles, doubles, and perturbative triples [CCSD(T)] levels
\citep{Rag89, ccreview, Shavitt09} and the correllation-consistent family of
basis sets in the cc-pVXZ from \citep{Dunning89, cc-pVXZ, aug-cc-pVXZ}.  The
cc-pV(X+d)Z basis sets are used for the phosphorus atom, but this distinction
should be understood by the reader to be implied for the remainder of the
discussion.  Additionally, restricted Hartree-Fock reference wavefunctions
\citep{ScheinerRHF87} and the PSI4 quantum chemistry software \citep{psi4} are
utilized except where noted.

The electronically excited states are computed vertically using the equation of
motion (EOM) formalism \citep{Stanton93EOM, Krylov07} of CCSD and are based on
the optimized CCSD(T)/aug-cc-pVTZ geometry.  Increasing the diffuseness of the
basis set allows one to observe the convergence of the excitation energy.
Early convergence in the series indicates a valence excited state while late
convergence is a sign that the excited state of interest is dipole-bound.  This
can them be corroborated by analysis of the orbitals involved from the most
diffuse basis set.  The double-zeta ($n$-aug-cc-pVDZ where $n=0-3$) and triple
zeta ($m$-aug-cc-pVTZ where $m=0-2$) basis sets are employed.  The vertical
electron binding energy (eBE), the energy required to removed the electron from
the system starting from the ground electronic state, is computed using the
ionization potential form of EOM: EOMIP-CCSD \citep{Stanton94:EOMIP}.

The rovibrational computations utilize the scheme developed previously
\citep{Huang08, Huang09, Huang11} that has met with much success as described
above.  The optimized CCSD(T)/aug-cc-pV5Z geometry is corrected for
core-electron inclusion by adding the differences in the bond lengths between
the CCSD(T) computations utilizing the Martin-Taylor (MT) core correlating
basis set \citep{Martin94} with and without inclusion of the core electrons:
$1s$ for carbon and $1s$, $2s$, and $2p$ for phosphorus.  From this reference
geometry, a quartic force field (QFF) is constructed.  A QFF is fourth-order
Taylor series expansion of the potential portion of the nuclear Hamiltonian
\citep{Fortenberry13Morse}.  It takes the form:
\begin{equation}
V=\frac{1}{2}\sum_{ij}F_{ij}\Delta_i\Delta_j +
\frac{1}{6}\sum_{ijk}F_{ijk}\Delta_i\Delta_j\Delta_k +
\frac{1}{24}\sum_{ijkl}F_{ijkl}\Delta_i\Delta_j\Delta_k\Delta_l, 
\label{VVib}
\end{equation} 
where $\Delta_i$ are displacements and $F_{ij\ldots}$ are the force constants.
The displacements are defined from C$_1$ as the terminal carbon atom opposite the
phosphorus:
\begin{align}
S_1(\Sigma^+) &= \mathrm{C}_1-\mathrm{C}_2\\
S_2(\Sigma^+) &= \mathrm{C}_2-\mathrm{C}_3\\
S_3(\Sigma^+) &= \mathrm{C}_3-\mathrm{P}\\
S_4(\Pi_{xz}) &= \angle\mathrm{C}_1-\mathrm{C}_2-\mathrm{C}_3-{\bf{y}}\\
S_5(\Pi_{xz}) &= \angle\mathrm{C}_2-\mathrm{C}_3-\mathrm{P}-{\bf{y}}\\
S_6(\Pi_{yz}) &= \angle\mathrm{C}_1-\mathrm{C}_2-\mathrm{C}_3-{\bf{x}}\\
S_7(\Pi_{yz}) &= \angle\mathrm{C}_2-\mathrm{C}_3-\mathrm{P}-{\bf{x}}.
\end{align}
A similar coordinate setup has been utilized previously on the
structurally-related $l$-C$_3$H$^+$ cation \citep{Huang13C3H+}.  The step sizes
for the bond length displacements are 0.005 \AA\ and 0.005 radians for the
bends.

625 total displacements are needed to describe the C$_3$P$^-$ QFF.  At each
displacement, seven quantum chemical computations are undertaken at the CCSD(T)
level.  Extrapolation of the aug-cc-pVTZ, aug-cc-pVQZ, and aug-cc-pV5Z energies
to the complete basis set (CBS) limit via a three-point formula \citep{Martin96}
provides the reference energy for each point.  Differences between the MT
computed energies including and not including core electrons are added to the
CBS energy.  Furthermore, Douglas-Kroll scalar relativistic corrections
\citep{Douglas74} available in the MOLPRO 2010.1 quantum chemistry program
\citep{MOLPRO} are also added to the CBS energy.  The CcCR QFF
\citep{Fortenberry11hoco} is created from this approach where the CBS energy
(``C''), core correlation (``cC''), and relativistic terms (``R'') provide for
a very detailed description of the molecular potential energy surface.

\begingroup
\begin{table*}

\caption{The C$_3$P$^-$ Vertical Excitation Energies (in eV), Ground State
eBEs$^a$ (in eV), and Oscillator Strengths.$^b$}

\begin{tabular}{l | c c c c | c | c | c c c}
\hline
\label{Vertical}

Transition & pVDZ & apVDZ & dapVDZ & tapVDZ & eBE & $f$ & pVTZ & apVTZ & dapVTZ
\\

\hline

2$\ ^1\Sigma^+ \leftarrow 1\ ^1\Sigma^+$ & 3.52 & 3.31 & 3.30 & 3.30 &      &
$2 \times10^{-4}$ & 3.35 & 3.23 & 3.24 \\
1$\ ^1\Pi \leftarrow 1\ ^1\Sigma^+$ & 8.23 & 5.04 & 3.98 & 3.81 & 3.80 &
$8 \times10^{-3}$ & 7.03 & 4.89 & 4.09 \\

\hline

\end{tabular}
\\
$^a$Computed with EOMIP-CCSD/t-aug-cc-pVDZ.\\
$^b$The $f$ values are for EOM-CCSD/d-aug-cc-pVTZ.\\
\end{table*}
\endgroup

The energy points are fitted to the force constants via a least-squares formula.
Refitting the surface provides the CcCR minimum equilibrium geometry as well as
the proper force constants including zero value gradients.  Transformation of
the force constants into Cartesian coordinates is accomplished via the INTDER
\citep{intder} program.  Rotational \citep{Mills72, Watson77} and vibrational
\citep{Papousek82} perturbation theory computations at second order (VPT2)
produce spectroscopic constants and anharmonic vibrational frequencies,
respectively, through the SPECTRO program \citep{spectro91}.  A $2\nu_3=\nu_2$
type-1 Fermi resonance is included in SPECTRO's VPT2 computations.
M{\o}ller-Plesset second-order electronic perturbation theory (MP2)
computations with the 6-31+G$^*$ basis set \citep{mp2, hehre72} in the
Gaussian09 program \citep{g09} provide the vibrational intensities.

\section{Results and Discussion}

\subsection{Vertical Electronically Excited States}

The ground electronic state C$_3$P has a $\tilde{X}\ ^2\Pi$ term and a
(core)$7\sigma^2 8\sigma^2 9\sigma^2 10\sigma^2 2\pi^4 11\sigma^2 3\pi^3$
electronic configuration \citep{delRio96}.  The anion's additional
electron occupies the $3\pi$ orbital to complete the shell and give a valence
$\tilde{X}\ ^1\Sigma^+$ ground state term.  The $3\pi$ highest occupied
molecular orbital (HOMO) is significant for the electronic properties of the
anion.  Computations of the first two vertically excited states of C$_3$P$^-$
are given in Table \ref{Vertical}.  Both excitations are out of the $3\pi$
HOMO.  The 2 $^1\Sigma^+$ state involves excitations into the $4\pi^*$ orbital,
while the 1 $^1\Pi$ state is produced by exciting the electron into a diffuse
$s$-type orbital.

The 1 $^1\Pi$ state is a candidate for classification as a dipole-bound excited
state.  The excitation energy decreases significantly as the spatial extent of
the orbitals increases, the excitations are, again, into high-diffuse orbitals
as is necessary for a dipole-bound state \citep{Skurski00, Morgan153}, and the
eBE and dipole-bound excitation energy are nearly coincident.  Since the
eBEs of dipole-bound states are known to be on the order of
$\sim 0.01$ eV \citep{Hammer03, Diken04}, the latter point is a clear marker of
dipole-bound behavior.  Additionally, the dipole moment of the C$_3$P
$\tilde{X}\ ^2\Pi$ state at the anion geometry is computed here by
CCSD(T)/aug-cc-pVTZ to be 4.20 D, in line with previous computations
\citep{Maclean08} and large enough to support a dipole-bound state.

The 2 $^1\Sigma^+$ state is actually valence.  Its excitation is not affected
by the increase in the diffuseness of the basis set and is well-below the eBE.
It is the product of a $\pi \rightarrow \pi^*$ excitation which is similar to
$\pi$ electron systems in polyunsaturated carbon chains, a commonly-used
application of the particle-in-a-box quantum mechanical model system.  Hence,
C$_5$P$^-$ and longer chains of this family will also likely possess a valence
excited state.  The C$_3$P$^-$ 2 $^1\Sigma^+$ excitation energy is well
converged to the 3.30 eV level for the double-zeta bases and similarly at 3.24
eV for the triple-zeta bases.

Even though the intensities for the transitions are very small, C$_3$P$^-$ is
still an excellent candidate for dipole-bound, gas phase synthesis starting
from the neutral-radical in the ISM or in circumstellar media.  Capture of the
electron by the neutral radical can proceed through the dipole-bound excited
state producing the valence, closed-shell anion.  The valence excited state
provides an additional avenue for energy dissipation.  This may likely increase
its interstellar lifetime giving an exceptionally high [C$_3$P$^-$]/[C$_3$P]
ratio, potentially greater than that observed for [C$_6$H$^-$]/[C$_6$H].  As a
result, C$_3$P$^-$ should be more readily observed than its corresponding
neutral radical since its electronic properties are more enhanced than the
other known interstellar anions.  In order to detect C$_3$P$^-$, information
regarding the $\tilde{X} ^1\Sigma^+$ valence ground electronic state's
rotational, vibrational, and rovibrational behavior is needed and provided here
for the first time from further quantum chemical computations.

\subsection{Rovibrational Analysis}

\begingroup
\begin{table*}

\caption{The CcCR C$_3$P$^-$ Force Constants Given in
mdyn/\AA$^n$$\cdot$rad$^m$)$^a$.}

\label{fc}

\centering

\begin{tabular}{c r c r c r c r c r}
\hline

F$_{11}$ & 11.566 227 & F$_{441}$ & -0.4428 & F$_{2222}$ & 219.70 & F$_{5432}$ &  -0.66 & F$_{7611}$ &   0.00 \\ 
F$_{21}$ &  0.596 174 & F$_{442}$ & -0.5107 & F$_{3111}$ &   0.15 & F$_{5433}$ &   0.23 & F$_{7621}$ &  -1.13 \\ 
F$_{22}$ &  7.708 659 & F$_{443}$ &  0.0197 & F$_{3211}$ &   0.12 & F$_{5444}$ &   0.57 & F$_{7622}$ &   0.77 \\ 
F$_{31}$ & -0.380 686 & F$_{541}$ & -0.0978 & F$_{3221}$ &  -0.38 & F$_{5511}$ &  -0.44 & F$_{7631}$ &  -0.50 \\ 
F$_{32}$ &  0.689 835 & F$_{542}$ &  0.0236 & F$_{3222}$ &   0.66 & F$_{5521}$ &   0.65 & F$_{7632}$ &  -0.19 \\ 
F$_{33}$ &  7.655 860 & F$_{543}$ & -0.0500 & F$_{3311}$ &  -0.93 & F$_{5522}$ &  -0.88 & F$_{7633}$ &   0.25 \\ 
F$_{44}$ &  0.209 473 & F$_{551}$ & -0.0499 & F$_{3321}$ &   0.31 & F$_{5531}$ &   0.53 & F$_{7644}$ &  -2.28 \\ 
F$_{54}$ & -0.010 817 & F$_{552}$ & -0.4968 & F$_{3322}$ &   0.11 & F$_{5532}$ &   1.80 & F$_{7654}$ &   3.47 \\ 
F$_{55}$ &  0.461 190 & F$_{553}$ & -0.6287 & F$_{3331}$ &   1.77 & F$_{5533}$ &   0.62 & F$_{7655}$ &   3.32 \\ 
F$_{66}$ &  0.209 484 & F$_{661}$ & -0.4424 & F$_{3332}$ &  -0.64 & F$_{5544}$ &   0.71 & F$_{7666}$ &   1.14 \\ 
F$_{76}$ & -0.010 826 & F$_{662}$ & -0.5109 & F$_{3333}$ & 177.78 & F$_{5554}$ &  -0.03 & F$_{7711}$ &  -0.97 \\ 
F$_{77}$ &  0.461 209 & F$_{663}$ &  0.0190 & F$_{4411}$ &   1.46 & F$_{5555}$ &   0.06 & F$_{7721}$ &   0.54 \\ 
F$_{111}$ &  -71.2065 & F$_{761}$ & -0.0952 & F$_{4421}$ &  -0.19 & F$_{6611}$ &   1.09 & F$_{7722}$ &  -1.29 \\ 
F$_{211}$ &   -0.2476 & F$_{762}$ &  0.0233 & F$_{4422}$ &   0.61 & F$_{6621}$ &  -0.42 & F$_{7731}$ &   0.65 \\ 
F$_{221}$ &   -2.4219 & F$_{763}$ & -0.0524 & F$_{4431}$ &   1.39 & F$_{6622}$ &   0.69 & F$_{7732}$ &   1.94 \\ 
F$_{222}$ &  -45.2600 & F$_{771}$ & -0.0492 & F$_{4432}$ &   0.77 & F$_{6631}$ &   1.53 & F$_{7733}$ &  -0.61 \\ 
F$_{311}$ &   -0.0455 & F$_{772}$ & -0.4972 & F$_{4433}$ &   0.07 & F$_{6632}$ &   1.31 & F$_{7744}$ &   0.03 \\ 
F$_{321}$ &    0.7443 & F$_{773}$ & -0.6289 & F$_{4444}$ &  -0.15 & F$_{6633}$ &  -0.38 & F$_{7754}$ &   2.40 \\ 
F$_{322}$ &   -1.7436 & F$_{1111}$ & 349.14 & F$_{5411}$ &  -0.05 & F$_{6644}$ &   1.89 & F$_{7755}$ &  -0.68 \\ 
F$_{331}$ &   -0.0806 & F$_{2111}$ &   0.82 & F$_{5421}$ &  -0.66 & F$_{6654}$ &  -1.66 & F$_{7766}$ &  -0.08 \\ 
F$_{332}$ &   -0.8594 & F$_{2211}$ &  -3.24 & F$_{5422}$ &  -0.07 & F$_{6655}$ &   0.71 & F$_{7776}$ &   0.00 \\ 
F$_{333}$ &  -41.3790 & F$_{2221}$ &   4.66 & F$_{5431}$ &   0.64 & F$_{6666}$ &  -0.25 & F$_{7777}$ &  -0.76 \\ 
        
\hline  
\end{tabular}
        
$^a$1 mdyn $=$ $10^{-8}$ N; The $n$ and $m$ values are exponents that
correspond to the number of units from the type of modes present in the
specific force constant.
        
\end{table*}
\endgroup
 
The fitting of the force constants needed to define the CcCR QFF and produce
the subsequent spectroscopic data of ground state $\tilde{X} ^1\Sigma^+$
C$_3$P$^-$ is very tight with a sum of residuals squared on the order of
$10^{-16}$ a.u.$^2$  The force constants for $l$-C$_3$P$^-$ are given in Table
\ref{fc} and follow the coordinate labeling given above.  For instance,
F$_{11}$ corresponds to the second-derivative of the C$_1-$C$_2$ bond.
Additionally, the harmonic, diagonal force constants are 
proportional to the bond strengths.  F$_{11}$ is the largest of these
indicating that the C$_1-$C$_2$ bond is the strongest.  This is corroborated by
the $R_{\alpha}$ vibrationally averaged bond lengths given in Table \ref{Data}.
The r$_0$(C$_1-$C$_2$) value is 1.259 055 \AA, the shortest bond length given
indicating that this is the strongest bond.  Interestingly, the C$_2-$C$_3$
bond strength from F$_{22}$ is roughly equal to the C$_3-$P bond strength from
F$_{33}$.  The phosphorus atom creates a longer bond length by default, but
these force constants and the bond lengths show that this molecule is neither
acetylenic nor cumulenic.  It should be thought of as an intermediate between
the two classifications with the C$_1-$C$_2$ bond more acetylenic and the rest
of the molecule as cumulenic.

\begingroup
\begin{table*}

\caption{The C$_3$P$^-$ and Single $^{13}$C Isotopologue CcCR Equilibrium and
Zero-Point ($R_{\alpha}$ vibrationally-averaged) Geometries, Rotational
Constants, Vibration-Rotation Interaction Constants, Quartic (D) and Sextic (H)
Distortion Constants, CCSD(T)/aug-cc-pVTZ Dipole Moment, and Vibrational
Frequencies and Intesnities$^a$.}

\label{Data}

\centering

\small

\begin{tabular}{l | r r r r r} 
\hline\hline

                  & C$_3$P$^-$   & $^{13}$CCCP$^-$ & C$^{13}$CCP$^-$ & CC$^{13}$CP$^-$  & MP2/6-31+G$^*$\\ 
\hline
r$_0$(C$_1-$C$_2$) (\AA) & 1.259 055 & 1.259 200 & 1.259 058 & 1.259 065 \\
r$_0$(C$_2-$C$_3$) (\AA) & 1.345 935 & 1.345 909 & 1.345 920 & 1.345 906 \\
r$_0$(C$_3-$P) (\AA)     & 1.581 669 & 1.581 612 & 1.581 683 & 1.581 652 \\
$B_0$         (MHz)      &  2798.0   &  2696.0   &  2769.6   &  2798.0\\
$\alpha^B$ 1  (MHz)      &    13.3   &    13.2   &  12.8     &   12.9 \\
$\alpha^B$ 2  (MHz)      &    12.6   &    11.8   &  12.4     &   12.4 \\
$\alpha^B$ 3  (MHz)      &     4.3   &     4.0   &   4.2     &    4.3 \\
$\alpha^B$ 4  (MHz)      &    -5.1   &    -4.9   &  -4.9     &   -5.0 \\
$\alpha^B$ 5  (MHz)      &    -7.9   &    -7.7   &  -7.6     &   -7.8 \\
\hline                                                    
r$_e$(C$_1-$C$_2$) (\AA) & 1.262 119 & --      & --       & --       \\
r$_e$(C$_2-$C$_3$) (\AA) & 1.343 872 & --      & --       & --       \\
r$_e$(C$_3-$P) (\AA)     & 1.580 317 & --      & --       & --       \\
$B_e$ (MHz)              &  2800.1   &  2698.0 &  2771.8  &  2800.1  \\
$D_e$ (kHz)              &  0.202    &  0.186  &  0.199   &  0.202    \\
$H_e$ ($\mu$Hz)          & 1.717     &  1.654  &  1.577   &  1.717    \\
$\mu_z$$^b$            &  6.22 D   & --      & --       & --       \\ 
\hline                                                    
$\omega_1(\sigma)$ Symm.~Carbon Stretch (cm$^{-1}$) & 1979.2 & 1959.0 & 1930.9 & 1971.8 & 1984.3 (483) \\
$\omega_2(\sigma)$ C$_2$ Shuttle (cm$^{-1}$)        & 1481.5 & 1470.5 & 1481.2 & 1439.3 & 1482.2 (93) \\
$\omega_3(\sigma)$ C$_3-$P stretch (cm$^{-1}$)      &  692.0 & 681.7  & 686.6  & 692.0  & 689.2 (7)\\
$\omega_4(\pi)$ Internal Bend (cm$^{-1}$)           &  479.9 & 479.4  & 472.5  & 469.6  & 477.0 (4)\\
$\omega_5(\pi)$ External Bend (cm$^{-1}$)           &  194.8 & 192.3  & 192.4  & 192.6  & 195.5 (10)\\
Harmonic Zero-Point (cm$^{-1}$)                     & 2751.0 & 2727.3 & 2714.3 & 2713.8 & 2750.4 \\
$\nu_1(\sigma)$ Symm.~Carbon Stretch (cm$^{-1}$) & 1939.7 & 1920.0 & 1913.8 & 1932.6 \\
$\nu_2(\sigma)$ C$_2$ Shuttle (cm$^{-1}$)        & 1456.5 & 1446.0 & 1456.0 & 1415.5 \\
$\nu_3(\sigma)$ C$_3-$P stretch (cm$^{-1}$)      &  684.0 &  674.5 &  678.2 &  683.9 \\
$\nu_4(\pi)$ Internal Bend (cm$^{-1}$)           &  470.8 &  470.4 &  463.3 &  461.3 \\
$\nu_5(\pi)$ External Bend (cm$^{-1}$)           &  181.1 &  178.8 &  179.0 &  179.4 \\
Zero-Point (cm$^{-1}$)                           & 2730.0 & 2706.7 & 2698.4 & 2693.8 \\
\hline                                                    
$B_1$ (MHz) & 2784.7 & 2682.8 & 2756.8 & 2785.1 \\
$B_2$ (MHz) & 2785.4 & 2684.2 & 2757.1 & 2785.6 \\
$B_3$ (MHz) & 2793.7 & 2692.0 & 2765.3 & 2793.8 \\
$B_4$ (MHz) & 2803.1 & 2700.9 & 2774.5 & 2803.0 \\
$B_5$ (MHz) & 2805.9 & 2703.7 & 2777.2 & 2805.8 \\

\hline
\end{tabular}

$^a$Given in parentheses in km/mol beside the MP2/6-31+G$^*$ harmonic
frequencies.\\
$^b$The coordinates used to generate the Born-Oppenheimer dipole moment
component (in \AA\ with the centre-of-mass at the origin) are: C$_1$, 0.000000,
0.000000, -2.629136; C$_2$, 0.000000, 0.000000, -1.367016; C$_3$, 0.000000,
0.000000, -0.023143; P, 0.000000, 0.000000, 1.557174\\

\end{table*}
\endgroup

In any case, the spectroscopic constants, dipole moment, and vibrational
frequencies are given in Table \ref{Data} along with the single $^{13}$C
substituted isotopologues.  Phosphorus has only one stable isotope, $^{31}$P.
C$_3$P$^-$ is strongly dipolar with a 6.22 D dipole moment as determined from
center-of-mass computations.  As a result, its rotational intensities should be
brighter than the neutral-radical giving further evidence that the anion may be
more easily observed in the ISM than the neutral.  The rotational constants are
not strongly affected by vibrational averaging with $B_0$ equal to 2798.0 MHz
while $B_e$ is only 2.1 MHz greater at 2800.1 MHz.  Substitution of the
terminal C$_1$ atom with $^{13}$C reduces the rotational constants by 102.0
MHz.  Substitution of the other carbon atoms is not as significant.  The
quartic (D) and sextic (H) distortion constants are exceptionally small for
this molecule and may only show up in very high-resolution experiments.  

While ground-based telescopes such as the Atacama Large
Millimeter/submillimeter (ALMA) array are excellent for observing molecular
rotational spectra, the growth of space-based or airborne infrared telescopes
such as the upcoming James Webb Space Telescope or the Stratospheric
Observatory for Infrared Spectroscopy (SOFIA), respectively, has enhanced
vibrational spectral analysis for astronomical applications.  The vibrational
frequencies of C$_3$P$^-$ are given in the middle of Table \ref{Data}.  The
descriptions of the fundamental vibrational modes are not as straightforward as
one would assume for a $C_{\infty v}$ molecule.  The first two modes are
defined here as the symmetric and antisymmetric C$-$C stretches for the three
carbon atom portion of the molecule.  The last two modes are the internal and
external bends.  The former mode is dominated by the C$_2$ and C$_3$ atoms
moving opposite of one another relative to the central axis.  The latter
bending vibrational motion is largely described by the C$_1$ and P atoms moving
in concert with one another relative to the central axis.  The carbon atom,
naturally, covers a greater distance in this mode due to its smaller relative
mass, but the two atoms are vibrating in concert.

The CcCR QFF frequencies are not strongly anharmonic.  While anharmonic
corrections are necessary for accurate descriptions of the vibrational modes,
they do not shift as greatly from the harmonic values as is typical.  Most of
this behavior is probably due to the increased mass of the system largely from
the phosphorus atom, but it is also from the tight bonding present in
C$_3$P$^-$.  Anions are not often thought be tightly bound due to the increased
electron-electron repulsion and likely increase in antibonding-character, but
C$_3$P$^-$ appears to be so bound from the data given.  In fact, the C$-$C
stretching frequencies in C$_3$P$^-$ are of the same order as their
counterparts in C$_3$H$^+$ \citep{Huang13C3H+}.  As a result, the vibrational
frequencies provided here should be very good representations of physical
reality.  The frequencies fit SOFIA's range very well, and the $\nu_1$
frequency at 1939.7 cm$^{-1}$ should be very bright since the double-harmonic
MP2 intensity is 483 km/mol.  It is also worth mentioning that the
MP2/6-31+G$^*$ harmonic frequency agree well with the CcCR harmonics.
MP2 is known to give Pauling points for some properties \citep{Sherrill09}, but
this correlation ($< 5.1$ cm$^{-1}$) is largely unprecedented but nicely
coincidental for the significantly lower-level MP2 computations. 

In the age of ALMA, unidentified lines (or ``U-lines'') are becoming even more
of an issue.  Many of these are likely to be caused by rotational transitions
of known molecules in vibrationally excited states.  As a result, the
vibrationally-averaged rotational constants of C$_3$P$^-$ are also provided at
the bottom of Table \ref{Data} for each fundamental mode.
Vibrationally-excited modes associated with the bending frequencies increase
the $B$ values, while the stretching vibrational modes decrease them. $^{13}$C
substitution largely affects the rotational constants in similar ways between
isotopologues as it does for $B_0$.  Finally, a limited line list for the $J
\leq 9$ rotational transitions of the $\nu=0$ and $\nu_5=1$ vibrational states
is provided in Table \ref{LLL} in order assist with further analysis.  Previous
work \citep{Fortenberry13C3H-} has shown that the rotational constants are
probably within 10 MHz of experiment putting the anticipated rotational
constants within $2J \times 10$ MHz for each line predicted in Table \ref{LLL},
and the relative intensities between the rotational transitions should be very
close to experiment.

\begingroup
\begin{table}

\caption{The C$_3$P$^-$ 150 K CcCR $\nu=0$ and $\nu_5=1$ Rotational Limited Line
List with $\Delta E_{J+1 \rightarrow J}, J \leq 9$ with Frequencies in MHz and
Intensities in MHz per molecule.}

\label{LLL}
\centering
\small
\begin{tabular}{l c c c c}
\hline\hline

Mode & J+1 & J & Frequency & 150K Intensity \\
$\nu=0$ & 1 & 0 &   5596.0 & 0.011 \\
$\nu=0$ & 2 & 1 & 11 192.0 & 0.042 \\
$\nu=0$ & 3 & 2 & 16 788.0 & 0.093 \\
$\nu=0$ & 4 & 3 & 22 384.0 & 0.164 \\
$\nu=0$ & 5 & 4 & 27 980.0 & 0.257 \\
$\nu=0$ & 6 & 5 & 33 575.9 & 0.367 \\
$\nu=0$ & 7 & 6 & 39 171.8 & 0.492 \\
$\nu=0$ & 8 & 7 & 44 767.7 & 0.628 \\
$\nu=0$ & 9 & 8 & 50 363.5 & 0.792 \\
$\nu=0$ & 10 & 9 & 55 959.3 & 0.961 \\
\hline
$\nu_5=1$ & 1 & 0 &   5611.8 & 0.002 \\
$\nu_5=1$ & 2 & 1 & 11 223.6 & 0.006 \\
$\nu_5=1$ & 3 & 2 & 16 835.4 & 0.014 \\
$\nu_5=1$ & 4 & 3 & 22 447.2 & 0.025 \\
$\nu_5=1$ & 5 & 4 & 28 058.9 & 0.040 \\
$\nu_5=1$ & 6 & 5 & 33 670.7 & 0.057 \\
$\nu_5=1$ & 7 & 6 & 39 282.4 & 0.076 \\
$\nu_5=1$ & 8 & 7 & 44 894.0 & 0.097 \\
$\nu_5=1$ & 9 & 8 & 50 505.7 & 0.122 \\
$\nu_5=1$ & 10 & 9 & 56 117.2 & 0.148 \\
\hline
\end{tabular}

\end{table}
\endgroup

\section{Conclusions}

The electronic properties of C$_3$P$^-$ are in line with those for the known
interstellar anions indicating that it may be a viable molecule waiting to be
detected in the ISM.  Attachment of an additional electron to the corresponding
neutral-radical via the anion's dipole-bound excited state is a likely creation
mechanism for this anion.  Additionally, C$_3$P$^-$ possesses not only a
valence ground electronic state but also a valence excited state.  The second
excited state should increase the molecule's lifetime.  Since
[C$_6$H$^-$]/[C$_6$H] $>1$ \citep{Agundez08}, it is likely that
[C$_3$P$^-$]/[C$_3$P] is as well since both radicals possess strongly dipolar
ground electronic states available for binding the extra electron leading to
the observed valence anion.  The valence excited state in C$_3$P$^-$ should
boost the [C$_3$P$^-$]/[C$_3$P] ratio further due to the enhanced stability in
the phosphorus anion.  Furthermore, the anion's dipole moment is greater than
that in the neutral-radical enhancing the anion's rotational signal.  As a
result, C$_3$P$^-$ may be a better candidate for detection than C$_3$P.

In order to assist in potential searches, the spectroscopic constants and
vibrational frequencies for C$_3$P$^-$ are provided in this work.  Ground state
C$_3$P$^-$ is tightly bound, especially considering that it is an anion, and
not strongly anharmonic.  Hence, the CcCR approach utilized previously with
very good results should be similarly accurate in this system.  The $\nu_1$
C$-$C stretch is very bright, and the vibration-rotation interaction constants
are small.  In summary, C$_3$P$^-$ is an excellent interstellar molecular
candidate, and the data provided in this work should assist in spectral
searches for this anion whether in the laboratory or in the ISM.

\section{Acknowledgements}

Georgia Southern University is acknowledged for the provision of start-up funds
necessary to complete this work.  Additionally, thanks go to Prof.~T.~Daniel
Crawford of Virginia Tech for the computer resources utilized as a part of this
work.


\bibliographystyle{mn2e}




\end{document}